# Quantum Crystallography[*] : Projectors & kernel subspaces preserving N-representability

Chérif F. Matta,[(1-4)] Lou Massa[(5,6)]*

[(1)] Department of Chemistry and Physics, Mount Saint Vincent University, Halifax, Nova Scotia, Canada B3M 2J6. [(2)] Dalhousie University, Halifax, Nova Scotia, Canada B3H 4J3. [(3)] Saint Mary's University, Halifax, Nova Scotia, Canada B3H 3C3. [(4)] Département de chimie, Université Laval, Québec, Québec, Canada G1V 0A6. [(5)] Department of Chemistry, Hunter College, City University of New York, New York, NY 10065, USA. [(6)] Departments of Chemistry and Physics, Graduate Center, City University of New York, New York, NY 10016, USA.

*E-mail: lmassa@hunter.cuny.edu

**Abstract**

Consider a projector matrix **P**, representing the first order reduced density matrix $\rho_{1(DET)}(\mathbf{r},\mathbf{r'}) = 2\operatorname{tr}\mathbf{P}\psi(\mathbf{r})\psi^{\dagger}(\mathbf{r'})$ in a basis of orthonormal atom-centric basis functions. A mathematical question arises, and that is, how to break **P** into its natural component kernel projector matrices, while preserving $N$-representability of $\rho_{1(DET)}$. The answer relies upon 2-projector triple products, $\mathbf{P'}_j\mathbf{P}\mathbf{P'}_j$. The triple product solutions, applicable within the quantum crystallography of large molecules, are determined by a new form of the Clinton equations, which - in their original form - have long been used to ensure $N$-representability of density matrices consistent with X-ray diffraction scattering factors.

---

1.	Introduction

The essential problem in the coherent X-ray diffraction experiment is that of atomic structure. One determines thereby atom positions and vibrations. For this purpose, the electron density represented as a sum of spherical atoms proved to be useful [1]. It was later realized that atoms described as carriers of multipolar densities would accurately represent the electron density in the bonding region between atomic nuclei [2], [3], [4], [5], [6], [7], [8], [9], [10], [11]. Such representations, while sufficient to solve the problem of atomic structure, are not mandated to be $N$-representable[12], [13], which is the principal problem addressed in this paper[14].

The availability of highly accurate experimental and theoretical electron densities has given rise to the strong synergy of X-ray crystallography on one side and quantum mechanics on the other [15], as evidenced by the emergence of the topological analyses of the electron density and related approaches epitomized by Bader's Quantum Theory of Atoms in Molecules (QTAIM) [16], [17], [18], [19], [20], [21], [22], [23].

The new field of quantum crystallography has taken a variety of different directions[24], [25]. Although one of the authors introduced the term "quantum crystallography" some time ago, the reader is encouraged to recognize the expanded recent use of that term aims at a field whose goal is not just that of obtaining density matrices or wavefunctions from experimental structure factors. The latter would be a too narrowly-focused definition which does not represent the way the term is intended by many scientists today. This topic is discussed in a recent review by Macchi [26]. Following on earlier work of Henderson and Zimmerman [27] and Gritsenko and Zhidomirov [28], the constraining of wavefunctions to experimental structure factors and to the energy variational principle has been pioneered by Jayatilaka *et al.*[29], [30], [31], [32].

Gas-phase quantum chemical calculation data is now being applied to construct databases for large systems using, for example, the extremely localized molecular orbitals (ELMO) method [33], [34], [35], [36]. It is expected that an X-ray constrained ELMO approach will soon be proposed in future works by those authors. Another approach is that of constraining density matrices to conform to the mathematical requirement of being $N$-representable [37], [38], [12], [39], which, when applied to the Kernel Energy Method (KEM) of molecular fragmentation [40], [41], [42], [43], [44], [45], [46], is here explored.

Quantum mechanical electronic structure must be *N*-representable to be consistent with the physical indistinguishability of electrons [38], [12]. It is well-known from the early work on quantum crystallography [40], [47], [48], [45] that single determinant wavefunction *N*-representability is ensured in the electron density matrix,

$$\rho_{1(DET)}(\mathbf{r},\mathbf{r}') = 2\,\text{tr}\,\mathbf{P}\boldsymbol{\psi}(\mathbf{r})\boldsymbol{\psi}^{\dagger}(\mathbf{r}'), \tag{1}$$

where $\boldsymbol{\psi}(\mathbf{r})\boldsymbol{\psi}^{\dagger}(\mathbf{r}')$ is a matrix of orthonormal basis products, where the factor 2 is associated with the double occupation of the ground state molecular orbitals, and **P** is a normalized projector [37], *i.e.*

$$\mathbf{P}^2 = \mathbf{P}, \tag{2}$$

and

$$\text{tr}\,\mathbf{P} = N. \tag{3}$$

The Clinton equations for extracting *N*-representable **P** from the X-ray experiment are of the form:

$$\mathbf{P}_{n+1} = 3\mathbf{P}_n^2 - 2\mathbf{P}_n^3 + \lambda_N^n \mathbf{1}_{N \times N}, \tag{4}$$

subject to constraints that **P** is normalized and the crystallographic *R*-factor is minimized. McWeeny derived the last equation in absence of the X-ray constraints the inclusion of which came later in the form of the Clinton equations. McWeeny's purpose was to "purify" an initial guess matrix into idempotent form [49].

In this article attention is primarily focused on molecular crystals since the eventual goal is the modeling of protein and nucleic acid crystals. The Clinton equations (4) assumes linearly independent crystal scattering data considerably greater in number than the number of independent elements in **P**. As the size of the molecules enveloped in the crystal unit cell grow sufficiently large, the number of elements in the matrix **P** can outweigh the number of independent X-ray scattering data. In such a case, we have suggested the use of multipole representations to solve the atomic structure, and quantum chemical calculations to solve the electronic structure [50], [13], [14]. Moreover, for such large molecular systems, in the Born-Oppenheimer approximation, one obtains an electronic solution in the form of a kernel energy method (KEM) calculation, *i.e.*, the molecular energy is written [51], [45]:

$$E_{\text{total} \atop (\text{KEM})} = \sum_{a=1}^{m-1} \sum_{b=a+1}^{m} E_{ab} - (m-2) \sum_{c=1}^{m} E_c , \qquad (5)$$

where $m$ is the number of single kernels, and where $E_{ab}$ is the energy of double kernel $ab$ and $E_c$ is the energy of single kernel $c$.

The density matrix follows an analogous KEM summation to the above (Eq. (5)), that is [52]:

$$\rho_{1(\text{DET})}(\mathbf{r},\mathbf{r}') = \sum_{a=1}^{m-1} \sum_{b=a+1}^{m} \rho_{1(ab)}(\mathbf{r},\mathbf{r}') - (m-2) \sum_{c=1}^{m} \rho_{1(c)}(\mathbf{r},\mathbf{r}') . \qquad (6)$$

We have shown that density matrices of the kernel fragments can be "summed" together to form a full matrix $\mathbf{P}_0$ representing the full molecule [53]. This $\mathbf{P}_0$, entered as a first iterant into the Clinton equations, delivers a projector $\mathbf{P}$ representing the full molecule accurately and with *N*-representability intact [53].

We now proceed to breaking the full molecule projector $\mathbf{P}$ into its natural kernel matrix components $\mathbf{P}'$ in such a way that *N*-representability is preserved.

## 2.  Results

Suppose we have a known projector operator $\hat{\mathbf{P}}$ onto a Hilbert space spanned by orthonormal atomic orbitals. In Dirac notation, the projector is displayed as:

$$\hat{\mathbf{P}} = \sum_{i=1}^{N} |\mathbf{P}_i\rangle\langle\mathbf{P}_i| . \qquad (7)$$

Similarly, consider a projector onto a subspace of $\hat{\mathbf{P}}$, also spanned by orthonormal atomic orbitals (AOs), let us call it

$$\hat{\mathbf{P}}' = \sum_{j=1}^{N'} |\mathbf{P}'_j\rangle\langle\mathbf{P}'_j| . \qquad (8)$$

The orbitals of $\hat{\mathbf{P}}$ and $\hat{\mathbf{P}}'$ are not necessarily defined as orthonormal to each other. We now have a sum of 2-projector triple products:

$$\sum_{j=1}^{N'} \hat{\mathbf{P}}'_j \hat{\mathbf{P}} \hat{\mathbf{P}}'_j = \sum_{j=1}^{N'} |\mathbf{P}'_j\rangle \left( \sum_{i=1}^{N} \langle\mathbf{P}'_j|\mathbf{P}_i\rangle\langle\mathbf{P}_i|\mathbf{P}'_j\rangle \right) \langle\mathbf{P}'_j| . \qquad (9)$$

Eq. (9) rearranged becomes:

$$\sum_{j=1}^{N'} \hat{\mathbf{P}}_j' \hat{\mathbf{P}} \hat{\mathbf{P}}_j' = \sum_{j=1}^{N'} |\mathbf{P}_j'\rangle\langle\mathbf{P}_j'| \left( \sum_{i=1}^{N} \langle\mathbf{P}_j'|\mathbf{P}_i\rangle^2 \right). \tag{10}$$

Eqs. (10) suggests re-writing in the form:

$$\sum_{j=1}^{N'} \hat{\mathbf{P}}_j' \hat{\mathbf{P}} \hat{\mathbf{P}}_j' = \sum_{j=1}^{N'} a_j^2 |\mathbf{P}_j'\rangle\langle\mathbf{P}_j'| \tag{11}$$

$$= \sum_{j=1}^{N'} a_j^2 \hat{\mathbf{P}}_j', \tag{12}$$

where:

$$a_j^2 \equiv \sum_{i=1}^{N} \langle\mathbf{P}_j'|\mathbf{P}_i\rangle^2, \quad \text{and} \quad \hat{\mathbf{P}}_j' \equiv |\mathbf{P}_j'\rangle\langle\mathbf{P}_j'|. \tag{13}$$

$\hat{\mathbf{P}}_j'$ is a projector onto a subspace of projector $\hat{\mathbf{P}}$, and $a_j^2$ is the probability that $\hat{\mathbf{P}}_j'$ belongs to the subspace of $\hat{\mathbf{P}}$. If

$$\hat{\mathbf{P}}' = \sum_{j=1}^{N'} \hat{\mathbf{P}}_j' \hat{\mathbf{P}} \hat{\mathbf{P}}_j', \tag{14}$$

then every $a_j^2 = 1$ and $\hat{\mathbf{P}}'$ belongs entirely to a subspace of the projector $\hat{\mathbf{P}}$.

Collecting in matrix form all the subspaces of **P** which correspond to molecular kernels, we obtain the kernel energy method (KEM) representation of the full molecule:

$$\mathbf{P}_{\text{total} \atop (\text{KEM})} = \sum_{a=1}^{m-1} \sum_{b=a+1}^{m} \mathbf{P}_{ab}' - (m-2) \sum_{c=1}^{m} \mathbf{P}_c'. \tag{15}$$

The question arises, given the projector **P** which carries the *N*-representability for the full molecule, how does one break **P** into its kernel projector subspaces **P**', indicated in the last equation. For this purpose, we employ the Clinton equations in a new form:

$$\mathbf{P}_{n+1}' = 3\left( \sum_{j=1}^{N'} \mathbf{P}_{j_n}' \mathbf{P} \mathbf{P}_{j_n}' \right)^2 - 2\left( \sum_{j=1}^{N'} \mathbf{P}_{j_n}' \mathbf{P} \mathbf{P}_{j_n}' \right)^3 + \lambda_{N'}^n \mathbf{1}_{N' \times N'}, \tag{16}$$

subject to the constraint:

$$\text{tr}\, \mathbf{P}' = N', \tag{17}$$

where $N'$ is the number of doubly-occupied orbitals associated with the kernel projector **P**', and where **P**' may represent either a single or a double kernel. The Lagrangian multiplier $\lambda$ is used to enforce the constraint (Eq. (17)). At every iteration, $\mathbf{P}'_{n+1}$ is diagonalized to obtain its eigenvalues and eigenfunctions. As convergence proceeds the non-zero eigenvalues will move towards 1, and the eigenfunctions allow reconstruction of new triple products, $\sum_{j=1}^{N'} \mathbf{P}'_{j_n} \mathbf{P} \mathbf{P}'_{j_n}$ for each new iteration.

The new form of the Clinton equations are used to extract from **P** all kernels belonging to kernel subspaces of **P**, thus, the KEM representation of **P** is determined and its *N*-representability is conserved.

In this case, the *N*-representable one-body density matrix in terms of the KEM expansion is:

$$\rho_{1(\text{DET})}(\mathbf{r},\mathbf{r}') = 2\,\text{tr}\left[\sum_{a=1}^{m-1}\sum_{b=a+1}^{m} \mathbf{P}'_{ab} - (m-2)\sum_{c=1}^{m} \mathbf{P}'_c\right]\psi(\mathbf{r})\psi^\dagger(\mathbf{r}'), \tag{18}$$

where the chemical identity of each kernel is preserved.

The X-ray scattering factors are obtained as the Fourier transform of the diagonal elements of $\rho_{1(\text{DET})}(\mathbf{r},\mathbf{r}')$. Using $f(\mathbf{K})$ as the Fourier transform of basis orbital products,

$$F(\mathbf{K}) = 2\,\text{tr}\left[\sum_{a=1}^{m-1}\sum_{b=a+1}^{m} \mathbf{P}'_{ab} - (m-2)\sum_{c=1}^{m} \mathbf{P}'_c\right]\mathbf{f}(\mathbf{K}). \tag{19}$$

The scattering factor contribution of each kernel is given as:

$$F'(\mathbf{K}) = 2\,\text{tr}\,\mathbf{P}'\mathbf{f}(\mathbf{K}). \tag{20}$$

The accuracy of the KEM representation may be judged according to the crystallographic *R*-factor, calculated as usual:

$$R-\text{factor} = \frac{\sum ||F_{\text{observed}}| - |F_{\text{calculated}}||}{\sum |F_{\text{observed}}|}. \tag{21}$$

Given the knowledge of $\rho_{1(\text{DET})}(\mathbf{r},\mathbf{r}')$ in KEM form, the diagonal elements of the spinless two-body density matrix is given as:

$$\rho_{2(\text{DET})}(\mathbf{r},\mathbf{r}') = \begin{vmatrix} \rho_{1(\text{DET})}(\mathbf{r}) & \tfrac{1}{2}\rho_{1(\text{DET})}(\mathbf{r},\mathbf{r}') \\ \rho_{1(\text{DET})}(\mathbf{r}',\mathbf{r}) & \rho_{1(\text{DET})}(\mathbf{r}') \end{vmatrix}, \tag{22}$$

which is also exactly *N*-representable [54], [55], yielding a calculation of the total energy of a molecule in which the density matrices are expanded in kernel fragments.

$$E = \int \hat{h}_1 \rho_{1(\text{DET})}(\mathbf{r},\mathbf{r}')\Big|_{\mathbf{r}'\to\mathbf{r}} d\mathbf{r} + \int \hat{h}_{12}\rho_{2(\text{DET})}(\mathbf{r},\mathbf{r}')\,d\mathbf{r}d\mathbf{r}', \tag{23}$$

where

$$\hat{H} = \hat{h}_1 + \hat{h}_{12} \tag{24}$$

where $\hat{h}_1$ and $\hat{h}_{12}$ are, respectively, the one- and two-body terms in the Hamiltonian $\hat{H}$.

## 3. Significance of the *N*-Representability of the Kernel Energy Method (KEM)

The kernel energy method (KEM) fragmentation approach is meant to "inject" quantum mechanics into a large macromolecular structure obtained from traditional crystallography. That is, first, crystallography delivers the atomic positions which are then "dressed up" by proper quantum mechanical density (matrices). How can this be done? By breaking the electronic problem into pieces, the calculations can be enormously sped-up on the grounds of (*a*) the exponential scaling of electronic structure calculations, but also on the grounds of (*b*) the ever-increasing availability of massively parallelized computational software and hardware, applied to the KEM formalism, which is inherently parallelizable.

The imposition of *N*-representability on the electron density (matrices) reconstructed from KEM is a prerequisite for extracting all quantum mechanical quantities from the appropriate operation of linear Hermitian operators. *N*-representability guarantees the consistency of the density (matrices) with an underlying antisymmetric wavefunction. By having these quantum mechanical density (matrices) we can study a range of properties such as for example, momentum densities or energies (molecular and/or atomic) that would otherwise be inaccessible [13], [56].

## 4. Conclusions

We have indicated the physical interpretation of the double-projector triple products. This delivers projectors onto the subspaces of full molecule *N*-representable **P**, weighted by the probability that subspace projectors belong to a subspace of **P**. If each kernel normalized projector **P'** belongs entirely to **P** then that allows the breakup of **P** into a summation of KEM projectors **P'.** The KEM summation of all **P'** projector matrices, which equal **P**, therefore, preserves the *N*-representability of the KEM expression of $\rho_{1(DET)}(\mathbf{r},\mathbf{r}')$. This allows an exactly *N*-representable $\rho_{1(DET)}(\mathbf{r},\mathbf{r}')$ to deliver a calculation of the full molecule structure factors in KEM form. Calculations of energy in its dependence upon kernels smaller than the full molecule will be numerically simpler than would otherwise occur using, directly, the density matrix of the full molecule.

We have recently shown how to combine kernels to obtain a KEM *N*-representable full molecule density matrix by using the Clinton equations [53]. Here, Eq. (9) is achieving the reverse strategy, that is, having a known full density matrix to break it down into kernel fragments that maintain its *N*-representablity. The first approach adopts a synthetic bottom-up philosophy while the present approach is analytical (Cartesian), that is, top-down. This is a novelty of the present paper.

**Acknowledgements**

The authors are indebted to Dr. Lulu Huang for discussions on quantum crystallography. Funding for this project was provided by the *U.S. Naval Research Laboratory* (project # 47203-00 01) and by a *PSC CUNY Award* (project # 63842-00 41) - (L. M.), the *Natural Sciences and Engineering Research Council of Canada* (NSERC), *Canada Foundation for Innovation* (CFI), and *Mount Saint Vincent University* - (C. F. M.).